\documentclass[pra,twocolumn]{revtex4-2}
\usepackage{amsthm}
\usepackage{amsmath}
\usepackage{latexsym}
\usepackage{amsfonts}
\usepackage{amssymb}
\usepackage{color}
\usepackage{bbm,dsfont}
\usepackage{graphicx}
\usepackage{hyperref}
\usepackage{subfigure}
\usepackage{caption}
\usepackage{xr}
\usepackage{mathrsfs}
\usepackage[noend]{algpseudocode}
\usepackage{algorithmicx,algorithm}
\usepackage{graphicx}
\usepackage{subfigure}
\usepackage{[number}
\usepackage{float}
\UseRawInputEncoding
\definecolor{gray}{RGB}{123,123,123}

\newtheorem{proposition?}{Proposition?}

\theoremstyle{definition}

\begin{document}
\title{Detection of Network and Genuine Network Quantum Steering}
\author{Zhihua Chen}
\thanks{Electronic address:  chenzhihua77@sina.com}
\author{Kai Wu}
\affiliation{School of Science, Jimei University, Xiamen 361021,China}
\author{Shao-Ming Fei}
\thanks{Electronic address:  feishm@cnu.edu.cn}
\affiliation{School of Mathematical Sciences, Capital Normal University, Beijing 100048, China}
\begin{abstract}
The quantum network correlations play significant roles in long distance quantum communication, quantum cryptography and distributed quantum computing. Generally it is very difficult to characterize the multipartite quantum network correlations such as nonlocality, entanglement and steering. In this paper, we propose the network and the genuine network quantum steering models from the aspect of probabilities in the star network configurations. Linear and nonlinear inequalities are derived to detect the network and genuine network quantum steering when the central party performs one fixed measurement. We show that our criteria can detect more quantum network steering than that from the violation of the $n$-locality quantum networks. Moreover, it is shown that biseparable assemblages can demonstrate genuine network steering in the star network configurations.
\end{abstract}

\maketitle

\section{Introduction}
Strong quantum correlations like quantum nonlocality in quantum networks can be established among distant parties sharing physical resources emitted by independent sources.
The quantum network is important in quantum information processing such as quantum cryptography, quantum communication and distributed quantum computation. Compared with the case of the parties sharing quantum entanglement from a single common source, it is much more difficult to characterize the correlations in quantum networks because of the non-convexity of the local network spaces. In quantum network scenarios, the parties hold the physical systems from different sources which are assumed to be independent with each other. Entanglement swapping is the simplest network scenario, where Alice, Bob and Charlie share two independent sources. When Bob performs the joint measurements on his two parties, Alice and Charlie can share the entanglement \cite{Zukowski}. In the bilocality scenario Alice and Charlie admit two independent hidden variables. It is then generalized to the $n$-locality scenarios.

Many efficient methods have been proposed to detect the quantum nonlocality in quantum networks, such as the explicit parametrization of network local models \cite{Branciard,Tavakoli1}, hierarchies of relaxations of the sets of compatible correlations \cite{Wolfe}, inflation technique \cite{inflation,Gisin}, network Bell inequalities \cite{nonlinear,Tavakoli2,Tavakoli3,170401,032119,020304} and numerical approaches \cite{neural, semidefinite}.

Motivated by the quantum network nonlocality, the network quantum steering has been proposed in \cite{Benjamin}, in which the intermediate parties are untrusted while the endpoints are trusted. Under the measurement performed on the intermediate parties, quantum network steering emerges when the sub-normalized state is entangled. Inequalities have been constructed to detect the network steering in the star network scenario with the central party trusted and all the edge parties untrusted \cite{Guangming}.
But

In this paper, we introduce the network steering and the genuine network steering from the central parties to the edge parties in star networks, from the aspect of the probability theory. Linear and nonlinear inequalities are constructed to verify the network steering and the genuine network steering for both linear and star networks when the central party performs the fixed measurement. These inequalities can detect more network steering than those of the network n-nonlocality. We demonstrate that the biseparable sub-normalized state is genuine network steerable by example.

\section{EPR-Steering}
Consider the bipartite EPR-steering scenario \cite{Wiseman,Pusey}. Alice and Bob share a bipartite quantum state $\rho_{AB}$. Alice performs a set of black-box measurements $\mathbf{x}$ on $\rho_{AB}$ with outcomes $\mathbf{a},$ denoted by $M_{\mathbf{x}}^{\mathbf{a}}$. The set of sub-normalized states $\{\tau_{\mathbf{x}}^{\mathbf{a}}\}_{\mathbf{x},\mathbf{a}}$ on Bob's side is called an assemblage. Each element in the assemblage is given by $\tau_{\mathbf{x}}^{\mathbf{a}}={\rm{Tr}}_A[(M_{\mathbf{x}}^{\mathbf{a}}\otimes \rm{I}_2)\rho_{AB}],$ where $\rm{I}_2$ is the identity matrix. Alice can not steer Bob if $\tau_{\mathbf{x}}^{\mathbf{a}}$ admits a local hidden state (LHS) model as follows,
\begin{eqnarray}
\begin{aligned}
\tau_{\mathbf{x}}^{\mathbf{a}}=\sum\limits_{\lambda}p(\lambda)p(\mathbf{a}|\mathbf{x},\lambda)\rho_{\lambda},
\end{aligned}
\end{eqnarray}
where $\lambda$ denotes the classical random variable distributed according to $p(\lambda)$ satisfying $\sum\limits_{\lambda}p(\lambda)=1,$
$\rho_{\lambda}$ is the hidden local state of Bob and $p(\mathbf{a}|\mathbf{x},\lambda)$ is the local response function of Alice.
If there are measurements such that $\tau_{\mathbf{x}}^{\mathbf{a}}$ does not admit an LHS model, $\rho_{AB}$ is said to be steerable from Alice to Bob.

The EPR-steering of $\rho_{AB}$ from Alice to Bob can also be described by the joint probability. Bob performs measurements $\mathbf{y}$ on the assemblage $\{\tau_{\mathbf{x}}^{\mathbf{a}}\}_{\mathbf{x},\mathbf{a}}$ with outcomes $\mathbf{b},$ denoted by $M_{\mathbf{y}}^{\mathbf{b}}$. The joint probabilities are $p(\mathbf{a},\textsl{b}|\textsl{x},\textsl{y})={\rm{Tr}}[M_{\textsl{y}}^{\textsl{b}}\sigma_{\textsl{x}}^{\textsl{a}}]$. $\rho_{AB}$ is steerable from Alice to Bob if there exist measurements $M_{\textsl{x}}^{\textsl{a}}$ and $M_{\textsl{y}}^{\textsl{b}}$ such that the joint probability does dot admit the following local hidden variable-local hidden state (LHV-LHS) model,
\begin{eqnarray}
\begin{aligned}
p(\textsl{a},\textsl{b}|\textsl{x},\textsl{y})=\sum\limits_{\lambda}p(\lambda)p(\textsl{a}|\textsl{x},\lambda)p_Q(\textsl{b}|\textsl{y},\rho_{\lambda}).
\end{aligned}
\end{eqnarray}

Consider the EPR-steering scenario of tripartite state $\rho_{ABC}$ shared by Alice, Bob and Charlie \cite{He,Cavalcanti,Costa,Jebaratnam}. Alice, Bob and Charlie perform measurements $\textsl{x},$ $\textsl{y}$ and $\textsl{z}$ with outcomes $\textsl{a},$ $\textsl{b}$ and $\textsl{c},$ denoted by $M_{\textsl{x}}^{\textsl{a}},$ $M_{\textsl{y}}^{\textsl{b}}$ and $M_{\textsl{z}}^{\textsl{c}}$, respectively. The joint probability is given by  $p(\textsl{a},\textsl{b},\textsl{c}|\textsl{x},\textsl{y},\textsl{z})=\rm{Tr}[\rho_{ABC}(M_{\textsl{x}}^{\textsl{a}}\otimes M_{\textsl{y}}^{\textsl{b}}\otimes M_{\textsl{z}}^{\textsl{c}})]$. $\rho_{ABC}$ is said to be tripartite steerable from Alice and Bob to Charlie if there are measurements $M_{\textsl{x}}^{\textsl{a}},$ $M_{\textsl{y}}^{\textsl{b}}$ and $M_{\textsl{z}}^{\textsl{c}}$ such that the joint probability  $p(\textsl{a},\textsl{b},\textsl{c}|\textsl{x},\textsl{y},\textsl{z})$
does not satisfy the fully local hidden variable and local hidden state model as follows,
\begin{eqnarray}
\begin{aligned}
&p(\textsl{a},\textsl{b},\textsl{c}|\textsl{x},\textsl{y},\textsl{z})\\
=&\sum\limits_{\lambda}p(\lambda)p(\textsl{a}|\textsl{x},\lambda)p(\textsl{b}|\textsl{y},\lambda)p_Q(\textsl{c}|\textsl{z},\rho^{C}_{\lambda}).
\end{aligned}
\end{eqnarray}

The state $\rho_{ABC}$ is tripartite steerable from Alice to Bob and Charlie  if there exist measurements $M_{\textsl{x}}^{\textsl{a}},$ $M_{\textsl{y}}^{\textsl{b}}$ and $M_{\textsl{z}}^{\textsl{c}}$ such that the joint probability  $p(\textsl{a},\textsl{b},\textsl{c}|\textsl{x},\textsl{y},\textsl{z})$
does not satisfy the local hidden variable and fully local hidden state model as follows,
\begin{eqnarray}
\begin{aligned}
&p(\textsl{a},\textsl{b},\textsl{c}|\textsl{x},\textsl{y},\textsl{z})\\
=&\sum\limits_{\lambda}p(\lambda)p(\textsl{a}|\textsl{x},\lambda)p_Q(\textsl{b}|\textsl{y},\rho^{B}_{\lambda})
p_Q(\textsl{c}|\textsl{z},\rho^{C}_{\lambda})\label{eq-full}.
\end{aligned}
\end{eqnarray}
where $\rho_{\lambda}^B$ and $\rho_{\lambda}^C$ are the local hidden states of Bob and Charlie, respectively.

The other tripartite steering of $\rho_{ABC}$ from Alice to Bob and Charlie is defined as in \cite{Costa}, for which there exist measurements $M_{\textsl{x}}^{\textsl{a}},$ $M_{\textsl{y}}^{\textsl{b}}$ and $M_{\textsl{z}}^{\textsl{c}}$ such that the joint probability  $p(\textsl{a},\textsl{b},\textsl{c}|\textsl{x},\textsl{y},\textsl{z})$
does not satisfy the local hidden variable and bipartite local hidden state model as follows,
\begin{eqnarray}
\begin{aligned}
&p(\textsl{a},\textsl{b},\textsl{c}|\textsl{x},\textsl{y},\textsl{z})\\
=&\sum\limits_{\lambda}p(\lambda)p(\textsl{a}|\textsl{x},\lambda)p_Q(\textsl{b},\textsl{c}|\textsl{y},\textsl{z},\rho^{BC}_{\lambda})\label{eq-bipartite},
\end{aligned}
\end{eqnarray}
where $\rho_{\lambda}^{BC}$ is the local hidden state of Bob and Charlie. In both scenarios given by Eq.(\ref{eq-full}) and Eq.(\ref{eq-bipartite}), a source sends
a classical message $\lambda$ to Alice with the probability $p(\lambda)$. The difference between (\ref{eq-full}) and (\ref{eq-bipartite}) is that the
corresponding local quantum states sent to Bob and Charlie are separable and entangled respectively.

\section{Network quantum steering}
Consider the networks that the $n$ parties are arranged in the star networks.
For the simplest scenario which has three parties and two sources,
the first two parties Alice and Bob$_1$ share the state $\rho_{AB_1}$ and the last two parties Bob$_2$ and Charlie share the state $\rho_{B_2C}.$ The intermediate party Bob, Bob$_1$ and Bob$_2$, performs the fixed measurement $\textsl{y}$ (without the input) with outcomes $\textsl{b}$,denoted as $M_{\textsl{y}}^{\textsl{b}}$,
The sub-normalized state under the measurement is
\begin{equation}
\sigma_{\textsl{b}}^{AC}={\rm{Tr}}_{B}[(\rm{I}_{\it{A}}\otimes M^{\textsl{b}}_{\textsl{y}}\otimes\rm{I}_{\textit{C}})(\rho_{\textit{AB}_1}
\otimes\rho_{\textit{B}_2\textit{C}})].
\end{equation}
$\sigma_b^{AC}$ admits a network local hidden state model (NLHS) if $\sigma_b^{AC}$ satisfies the following condition,
\begin{eqnarray}
\begin{aligned}
\sigma_{\textsl{b}}^{AC}=\sum\limits_{\lambda_1,\lambda_2}p(\lambda_1)
p(\lambda_2)p(\textsl{b}|\lambda_1,\lambda_2)\rho_{\lambda_1}^A\otimes\rho_{\lambda_2}^C.
\end{aligned}
\end{eqnarray}
The network state $\rho_{ABC}=\rho_{AB_1}\otimes\rho_{B_2C}$ demonstrates the network steering from the central party to the endpoint parties if there exists a fixed measurement $M_{\textsl{y}}^{\textsl{b}}$ such that $\sigma_{\textsl{b}}^{AC}$ does not admit NLHS \cite{Benjamin}.


The entanglement of $\sigma_{\textsl{b}}^{AC}$ can rule out the NLHS model from Bob to Alice and Charlie, but entanglement detection is a difficult problem, especially for
high dimensional quantum states. In \cite{Benjamin}, the authors only investigated the network steering scenarios with respect to some special states such as separable and unsteerable ones. To investigate the network steering for general quantum states, we define the network steering from the aspect of joint probabilities. We derive inequalities to detect if the sub-normalized states under the measurement performed by the central party admit the NLHS model, thus detecting the network steering.

Bob performs a fixed measurement $\textsl{y}$ with outcome $\textsl{b}$. The endpoint parties Alice and Charlie perform the measurements $\textsl{x}$ and $\textsl{z}$ on the sub-normalized state $\sigma_{\textsl{b}}^{AC}$ with outcomes $\textsl{a}$ and $\textsl{c}$. The joint probability $p(\textsl{a},\textsl{b},\textsl{c}|\textsl{x},\textsl{z})={\rm{Tr}}[(M_{\textsl{x}}^{\textsl{a}}\otimes M_{\textsl{y}}^{\textsl{b}})\sigma_{\textsl{b}}^{\textit{AC}}]$ admits a network local hidden variable and local hidden state model (NLHV-LHS) from
the central party Bob to the endpoint parties Alice and Charlie if the joint probability satisfies the following condition,
\begin{eqnarray}
\begin{aligned}
&p(\textsl{a},\textsl{b},\textsl{c}|\textsl{x},\textsl{z})\\
=&\sum\limits_{\lambda_1,\lambda_2} p(\lambda_1)p(\lambda_2)p_Q(\textsl{a}|\textsl{x},\rho_{\lambda_1}^A)\\
&\times p(\textsl{b}|\lambda_1,\lambda_2)p_Q(\textsl{c}|\textsl{z},\rho_{\lambda_2}^C),
\end{aligned}
\end{eqnarray}
where $p_Q(\textsl{a}|\textsl{x},\rho_{\lambda_1}^A)$ ($p_Q(\textsl{c}|\textsl{z},\rho_{\lambda_2}^C)$) is the probability generated from Alice's (Charlie's) system, $p(\textsl{b}|\lambda_1,\lambda_2)$ is the probability from Bob's system $B_1B_2$.

Consider the network composed of three parties Alice, Bob and Charlie and two resources $\lambda_1$ and $\lambda_2$. Alice and Charlie perform the mutually unbiased measurements $x^i$ with outcomes $a$ and $z^i$ with outcomes $c$ $(i=1,2,\,a,c=0,1)$ and Bob performs the fixed measurement $y=\{G_{00},G_{01},G_{10},G_{11}\}$  with four possible outcomes $b\equiv\{b_1,b_2\}\in\{0,1\}.$
Denote
$y_1=G_{00}-G_{11}-(G_{01}-G_{10}),$ $y_2=G_{00}-G_{11}+G_{01}-G_{10}$ and $y_3=G_{00}+G_{11}-G_{01}-G_{10}$. Then
\begin{eqnarray}
\begin{aligned}
&\langle x^i\otimes y^j\otimes z^k\rangle=\sum\limits_{a,b,c}(-1)^{a+t\cdot b+c}p(a,b,c|x^i,z^k),\\ \nonumber
\end{aligned}
\end{eqnarray}
with $\{b_1,b_2\}$ the string of 2 bits representing $b=0,1,2$ and $3,$ $t$ the string of 2 bits representing $j=1,2,3.$ We have the following Theorem, see proof in Appendix.

{\bf{Theorem 1.}} For line network, the probabilities that admit the NLHS-LHV model from Bob to Alice and Charlie satisfy the following inequalities,
\begin{eqnarray}\label{s-1}
\sum_{i=1}^3 |\langle x^i\otimes y^i\otimes z^i\rangle|\leq 1.
\end{eqnarray}

For the case of star networks, let us consider a star network composed of a central party $B=B_1\cdots B_n$ and $n$ edge parties $A_i$ shared by Bob and Alice$_i$, $i=1,\cdots,n$. The central party is separately connected to the $n$ edge parties. The edge parties perform the measurements $\bigotimes\limits_{k=1}^n M_{\textsl{x}_k}^{\textsl{a}_k}$ and the central party performs the fixed measurement $M_{\textsl{y}}^{\textsl{b}}$. The quantum state $\rho_{AB}=\bigotimes\limits_{k=1}^n\rho_{A_kB_k}$ admits an NLHV-LHS model from the central party to the edge parties if the joint probability satisfies
\begin{eqnarray}
\begin{aligned}
&p(\textsl{a}_1,\cdots,\textsl{a}_n,\textsl{b}|\textsl{x}_1,\cdots,\textsl{x}_n)=\rm{Tr}[(\bigotimes\limits_{k=1}^n M_{\textsl{x}_{\it{k}}}^{\textsl{a}_{\it{k}}}\otimes M_{\textsl{y}}^{\textsl{b}})\bigotimes\limits_{k=1}^n\rho_{\it{A}_{\it{k}}\it{B}_{\it{k}}}]\\ \nonumber
&=\sum\limits_{\lambda_k}p(\lambda_1)\cdots p(\lambda_n)p_Q(\textsl{a}_1|\textsl{x}_1,\rho_{\lambda_1}^{A_1})...p_Q(\textsl{a}_\textsl{n}|\textsl{x}_n,\rho_{\lambda_n}^{A_n})\\ \nonumber
&\times p(\textsl{b}|\lambda_1,\cdots,\lambda_n)
\end{aligned}
\end{eqnarray}
with  $\textsl{b}=\{\textsl{b}_1\cdots \textsl{b}_n\}$  and $p_Q(\textsl{a}_k|\textsl{x}_k,\lambda_k)=\rm{Tr}[M_{\textsl{x}_{\it{k}}}^{\textsl{a}_{\it{k}}}
\rho_{\lambda_k}^{\it{A}_{\it{k}}}].$

Let the edge parties perform the mutually unbiased measurements $x_k^{i_k}(k=1,\cdots,n)$ and  the fixed measurement performed by Bob is given by $\textsl{y}=\{G_{0\cdots0},G_{0\cdots 01},\cdots,G_{1\cdots1}\}$ with $2^n$ possible outcomes $\textsl{b}\equiv \{b_1,\cdots, b_{n}\}\in\{0,1\}^n$.  Set
\begin{equation}
y^{i_1\cdots i_n}=\left\{
\begin{aligned}
&\sum\limits_{t_1=0,t_2\cdots t_n}(-1)^{I\cdot T}(G_{t_1\cdots t_n}-G_{\bar{t}_1\cdots \bar{t}_n}),\\ \nonumber
&\hspace{5cm}i_1\cdots i_n\in C\\
&\sum\limits_{t_1=0,t_2\cdots t_n}(-1)^{I_0\cdot T}(G_{t_1\cdots t_n}+G_{\bar{t}_1\cdots \bar{t}_n}),\\ \nonumber
&\hspace{5cm}i_1\cdots i_n\in C'
\end{aligned}\right.
\end{equation}
where $I=\{i_1-1,\cdots,i_n-1\},$ $T=\{t_1,\cdots, t_n\}$, $``\cdot"$ represents the inner product, $I_0=\{i_1,\cdots,i_n\},$
$y_1^1=y^{1\cdots 1},$ $y_1^2=y^{2\cdots 2}$ and $y_1^3=y^{3\cdots 3},$
$C=\{111\cdots1,11\cdots122,\cdots, 221\cdots1,\cdots 22221\cdots1,\cdots\}$
(each string has either zero or even number of 2) and $C'=\{330\cdots0,00\cdots033,\cdots, 330\cdots0,\cdots 33330\cdots0,\cdots\}$
(each string has even number of 3).
Then we have
\begin{eqnarray}
\begin{aligned}
&\langle x^{i_1}_1\otimes x^{i_2}_2\otimes \cdots \otimes x^{i_n}_n\otimes y^{i_1\cdots i_n}\rangle\\ \nonumber
=&\sum\limits_{a_k^{i_k},\textsl{b}}(-1)^{\sum\limits_{k}a_k^{i_k}+\Re}p(a_1^{i_1},\cdots,a_n^{i_n},\textsl{b}|x_1^{i_1},\cdots,x_n^{i_n}),\\ \nonumber
\end{aligned}
\end{eqnarray}
where $\Re=I\cdot\Delta+b_1$ when $i_1\cdots i_n\in C,$ and  $\Re=I_0\cdot\Delta$ when $i_1\cdots i_n\in C',$
$\Delta=\{b_1,\cdots, b_n\}$ the string of 2 bits representing $b=0,1,\cdots,2^n-1$
when $b_1=0$ and $\Delta=\{\bar{b}_1,\cdots, \bar{b}_n\}$ when $b_1=1.$

We have the following conclusions, see proof in Appendix.

{\bf{Theorem 2.}} {\bf{(a)}} If $\rho_{AB}=\bigotimes\limits_{k=1}^n\rho_{A_{k}B_{k}}$
admits NLHV-LHS from the central party to the edge parties when Bob performs the fixed measurement in the two measurement settings, we have\\
(1) When $n$ is an odd number,
\begin{eqnarray}
\begin{aligned}
\label{2-set-nodd-1}
&\sum\limits_{i_1,\cdots,i_n\in C}|\langle x^{i_1}_1\otimes x^{i_2}_2\otimes \cdots \otimes x^{i_n}_n\otimes y^{i_1\cdots i_n}\rangle|^{\frac{2}{n}}\\
&\leq 2^{n-2}
\end{aligned}
\end{eqnarray}
and
\begin{eqnarray}
\begin{aligned}
\label{2-set-nodd-2}
&\sum\limits_{i_1,\cdots,i_n\in C}|\langle x^{i_1}_1\otimes x^{i_2}_2\otimes \cdots \otimes x^{i_n}_n\otimes y^{i_1\cdots i_n}\rangle|^{\frac{1}{n}}\\
&\leq 2^{n-2}\sqrt{2},
\end{aligned}
\end{eqnarray}

(2) When $n$ is an even number,
\begin{eqnarray}
\begin{aligned}
\label{2-set-neven-1}
&|\langle x^{1}_1\otimes x^{1}_2\otimes \cdots \otimes x^{1}_n\otimes y_1^1\rangle|^{\frac{2}{n}}\\
+&|\langle x^{2}_1\otimes x^{2}_2\otimes \cdots \otimes x^{2}_n\otimes y_1^2\rangle|^{\frac{2}{n}}\leq 1.
\end{aligned}
\end{eqnarray}
and
\begin{eqnarray}
\begin{aligned}
\label{2-set-neven-2}
&|\langle x^{1}_1\otimes x^{1}_2\otimes \cdots \otimes x^{1}_n\otimes y_1^1\rangle|^{\frac{1}{n}}\\
+&|\langle x^{2}_1\otimes x^{2}_2\otimes \cdots \otimes x^{2}_n\otimes y_1^2\rangle|^{\frac{1}{n}}\leq \sqrt{2}.
\end{aligned}
\end{eqnarray}

{\bf{(b)}} If $\rho_{AB}=\bigotimes\limits_{k=1}^n\rho_{A_{k}B_{k}}$ admits NLHV-LHS from the central party to the edge parties when Bob performs the fixed measurement in three measurement settings, we have\\
(1) When $n$ is an odd number,
\begin{eqnarray}
\begin{aligned}
\label{3-set-nodd-1}
&\sum\limits_{i_1,\cdots,i_n\in C\cup C'}|\langle x^{i_1}_1\otimes x^{i_2}_2\otimes \cdots \otimes x^{i_n}_n\otimes y^{i_1\cdots i_n}\rangle|^{\frac{2}{n}}\\
&\leq 2^{n-2}+2^{n-2}-1
\end{aligned}
\end{eqnarray}
and
\begin{eqnarray}
\begin{aligned}
\label{3-set-nodd-2}
&\sum\limits_{i_1,\cdots,i_n\in C\cup C'}|\langle x^{i_1}_1\otimes x^{i_2}_2\otimes \cdots \otimes x^{i_n}_n\otimes y^{i_1\cdots i_n}\rangle|^{\frac{1}{n}}\\
&\leq 2^{n-2}\sqrt{3}+2^{n-2}-1.
\end{aligned}
\end{eqnarray}
(2) When $n$ is an even number,
\begin{eqnarray}
\begin{aligned}
\label{3-set-neven-1}
&|\langle x^{1}_1\otimes x^{1}_2\otimes \cdots \otimes x^{1}_n\otimes y_1^1\rangle|^{\frac{2}{n}}\\
+&|\langle x^{2}_1\otimes x^{2}_2\otimes \cdots \otimes x^{2}_n\otimes y_1^2\rangle|^{\frac{2}{n}}\\
+&|\langle x^{3}_1\otimes x^{3}_2\otimes \cdots \otimes x^{3}_n\otimes y_1^3\rangle|^{\frac{2}{n}}\leq 1
\end{aligned}
\end{eqnarray}
and
\begin{eqnarray}
\begin{aligned}
\label{3-set-neven-2}
&|\langle x^{1}_1\otimes x^{1}_2\otimes \cdots \otimes x^{1}_n\otimes y_1^1\rangle|^{\frac{1}{n}}\\
+&|\langle x^{2}_1\otimes x^{2}_2\otimes \cdots \otimes x^{2}_n\otimes y_1^2\rangle|^{\frac{1}{n}}\\
+&|\langle x^{3}_1\otimes x^{3}_2\otimes \cdots \otimes x^{3}_n\otimes y_1^3\rangle|^{\frac{1}{n}}\leq \sqrt{3},
\end{aligned}
\end{eqnarray}
where $x_k^{i_k}$ $(i_k=1,2,3)$ are all mutually unbiased measurements for $k=1,\cdots,n,$
$x_1^0=x_2^0=\cdots x_n^0=\rm{I}_2$ are the identity matrices.

In the above studies we have concerned the sub-normalized state
under one fixed measurement performed by the central party. When the central party performs four measurements $B_i$ $(i=1,\cdots,4)$, and the edge parties performs three settings of measurements, we have the following, see proof in Appendix.

{\bf{Theorem 3.}} The star network state $\rho_{A_1\cdots A_n B}$ admits NLHS model from the central party to the edge parties if
\begin{eqnarray}
\begin{aligned}
|J_1|^{\frac{1}{n}}+|J_2|^{\frac{1}{n}}+|J_3|^{\frac{1}{n}}+|J_4|^{\frac{1}{n}}\leq 4,
\end{aligned}
\end{eqnarray}
where
\begin{eqnarray}
\begin{aligned}
&J_1=\langle\bigotimes\limits_{i=1}^n(x_i^1+x_i^2+x_i^3)\otimes B_1\rangle, \nonumber\\
&J_2=\langle\bigotimes\limits_{i=1}^n(x_i^1+x_i^2-x_i^3)\otimes B_2\rangle,\nonumber\\
&J_3=\langle\bigotimes\limits_{i=1}^n(x_i^1-x_i^2+x_i^3)\otimes B_3\rangle, \nonumber\\
&J_4=\langle\bigotimes\limits_{i=1}^n(-x_i^1+x_i^2+x_i^3)\otimes B_4\rangle \nonumber\\
\end{aligned}
\end{eqnarray}
and $x_i^{j}$ $(i=1,\cdots, n,j=1,2,3)$ are all mutually unbiased measurements.

As an example let us consider the star network $\rho_{A_1\cdots A_n B}=\bigotimes\limits_{i=1}^n \rho_{A_kB_k}$, where $\rho_{A_kB_k}=\frac{1-p}{4}\rm{I}_4+p |\phi\rangle\langle\phi|$ with
$|\phi\rangle=\frac{1}{\sqrt{2}}(|00\rangle+|11\rangle).$
Let $B_1=\bigotimes\limits_{i=1}^n(x_i^1+x_i^2+x_i^3),$ $B_2=\bigotimes\limits_{i=1}^n(x_i^1+x_i^2-x_i^3),$  $B_3=\bigotimes\limits_{i=1}^n(x_i^1-x_i^2+x_i^3)$ and $B_4=-x_i^1+x_i^2+x_i^3.$ From the theorem 3, we have that $\rho_{A_1\cdots A_n B}$ demonstrates the network steering when $p>\frac{\sqrt{3}}{3}.$ However, it has been shown that the state demonstrates the network nonlocality for $p>\frac{\sqrt{3}}{2}$ when the edge parties perform three settings of measurements \cite{Pan}.

In the following examples, we set $x_k^{i_k}$  to be pauli matrices, $G_{t_1\cdots t_n}=|\psi^+_{t_1\cdots t_n}\rangle\langle\psi^+_{t_1\cdots t_n}|$ and $G_{\bar{t}_1\cdots \bar{t}_n}=|\psi^-_{t_1\cdots t_n}\rangle\langle\psi^-_{t_1\cdots t_n}|$, where
$$|\psi^+_{t_1\cdots t_n}\rangle=\frac{1}{\sqrt{2}}(|t_1\cdots t_n\rangle+|\bar{t}_1\cdots \bar{t}_n\rangle),$$
$$|\psi^-_{t_1\cdots t_n}\rangle=\frac{1}{\sqrt{2}}(|t_1\cdots t_n\rangle-|\bar{t}_1\cdots \bar{t}_n\rangle)$$
with $\bar{t}$ representing the bit flip of $t,$ $t_1=0$ and $t_i\in\{0,1\}$ for $i\geq 2$.
For the special case of $n=2,$ we have
$G_{00}=|\psi_{00}^+\rangle\langle \psi_{00}^+|,$ $G_{11}=|\psi_{00}^-\rangle\langle \psi_{00}^-|,$ $G_{01}=|\psi_{01}^+\rangle\langle \psi_{01}^+|$ and $G_{10}=|\psi_{01}^-\rangle\langle \psi_{01}^-|,$ where $|\psi_{00}^{\pm}\rangle=\frac{1}{\sqrt{2}}(|00\rangle\pm|11\rangle)$ and $|\psi_{01}^{\pm}\rangle=\frac{1}{\sqrt{2}}(|01\rangle\pm |10\rangle).$

As another example let us consider the star network $\rho_{A_1\cdots A_n B}=\rho_{A_1B_1}\otimes \cdots\otimes\rho_{A_nB_n}$ with $\rho_{A_iB_i}=\frac{1}{4}(\rm{I}_4+c_1\sigma_1\otimes\sigma_1+c_2\sigma_2\otimes\sigma_2
+c_3\sigma_3\otimes\sigma_3)$ and $B=B_1B_2\cdots B_n$.
From (\ref{2-set-neven-1}),
$\rho_{A_1\cdots A_n B}$ is steerable from the central party to the edge parties when $\max\{c_1^2+c_2^2, c_1^2+c_3^2, c_2^2+c_3^1\}>1$ in two measurement settings (which is the same as the result in \cite{Kundu}), and from (\ref{3-set-neven-1}) $\rho_{A_1\cdots A_n B}$ is steerable when $c_1^2+c_2^2+c_3^2>1$ in three measurement settings if $n$ is even. When $n$ is odd, from (\ref{2-set-nodd-1}) and (\ref{2-set-nodd-2}) $\rho_{A_1\cdots A_n B}$ is steerable from the central party to the edge parties when
$$\sum\limits_{i_1\cdots i_n\in C}\sqrt[n]{(c_{i_1}\cdots c_{i_n})^2}>2^{n-2}$$
or
$$\sum\limits_{i_1\cdots i_n\in C}\sqrt[n]{|c_{i_1}\cdots c_{i_n}|}>2^{n-2}\sqrt{2}$$
in two measurement settings. From (\ref{3-set-nodd-1}) and (\ref{3-set-nodd-2}), $\rho_{A_1\cdots A_n B}$ is steerable when
$$\sum\limits_{i_1\cdots i_n\in C\bigcup C'}\sqrt[n]{(c_{i_1}\cdots c_{i_n})^2}>2^{n-1}-1$$
or
$$\sum\limits_{i_1\cdots i_n\in C\bigcup C'}\sqrt[n]{|c_{i_1}\cdots c_{i_n}|}>2^{n-2}\sqrt{3}+2^{n-2}-1$$
in three measurement settings with $c_0=1$.

In particular, when $|c_1|=|c_2|=|c_3|=p,$ we have that $\rho_{A_1\cdots A_n B}$ is steerable for $p>\frac{1}{\sqrt{2}}$ in two measurement settings, and for $p>\frac{1}{\sqrt{3}}$ (even $n$) or $p>p_0$ (odd $n$) in three measurement settings, where $p_0$ is the solutions of $2^{n-1}p+C_n^{n-1}p^{\frac{n-1}{n}}+\cdots+C_n^2p^{\frac{2}{n}}=2^{n-2}\sqrt{3}+2^{n-2}-1$
($p_0=0.589$ for $n=3$). Nevertheless, it has been shown in \cite{Pan} that $\rho_{A_1\cdots A_nB}$ is not of n-locality when $p>\frac{\sqrt{3}}{2}$ in three measurement settings and when $p>0.741$ in four measurement settings.

The general two-qubit quantum state $\rho_{A_iB_i}=\frac{1}{4}(\rm{I}_4+\sum\limits_j (a_j\sigma_j\otimes \rm{I}_2+ b_j\rm{I}_2\otimes \sigma_j)+\sum\limits_{kl}c_{kl}\sigma_k\otimes \sigma_l)$
is locally unitary equivalent to the state $\rho'_{A_iB_i}=\frac{1}{4}(\rm{I}_4+\sum\limits_j (a'_j\sigma'_j\otimes \rm{I}_2+ b'_j\rm{I}_2\otimes \sigma'_j)+\sum\limits_{k}d_{k}\sigma'_k\otimes \sigma'_k)$, where
$C=[c_{kl}]=U.D.V^{*}$ is the singular value decomposition of $C$ and $D=\rm{Diag}\{d_1,d_2,d_3\}$. Then $\rho_{A_1\cdots A_nB}$ is steerable from the central party to the edge parties when $\max\{d^2_1+d^2_2,d_1^2+d_3^2,d_2^2+d_3^2\}>1$ in two measurement settings, and when $d_1^2+d_2^2+d_3^2=\rm{Tr}[CC^T]>1$ in three measurement settings for even $n$. For odd $n$ $\rho_{A_1\cdots A_nB}$ is steerable from the central party to the edge parties when $$\sum\limits_{i_1\cdots i_n\in C}\sqrt[n]{|d_{i_1}\cdots d_{i_n}|^2}>2^{n-2}$$
or
$$\sum\limits_{i_1\cdots i_n\in C}\sqrt[n]{|d_{i_1}\cdots d_{i_n}|}>2^{n-2}\sqrt{2}$$ in two measurement settings, and when
$$\sum\limits_{i_1\cdots i_n\in C\bigcup C'}\sqrt[n]{(d_{i_1}\cdots d_{i_n})^2}>2^{n-1}-1$$
or $$\sum\limits_{i_1\cdots i_n\in C\bigcup C'}\sqrt[n]{|d_{i_1}\cdots d_{i_n}|}>2^{n-2}\sqrt{3}+2^{n-2}-1$$
in three measurement settings.

The inequalities in Theorem 1 and Theorem 2 can be used to detect the network steering when the central party performs a single unknown measurement in the star network scenarios composed of bell-diagonal states and general two-qubit states. Compared with the violation of $n$-locality, the violations of our inequalities detect more network steering for the star network scenarios composed of Werner states.

\section{Genuine Network Quantum steering}
Consider the star network with state $\rho_{\it{A_1A_2A_3B}}$, where Bob performs one fixed joint measurement $M_{\textsl{y}}^{\textsl{b}}$.
Entanglement can be generated among the edge parties when the central party performs the joint measurement. Here we consider if the network assemblage  $\{\sigma_b^{\it{A_1A_2A_3}}\}$ can be composed of the bi-separable local hidden states with $\sigma_{\textsl{b}}^{\it{A_1A_2A_3}}=\rm{Tr}_B[(\rm{I}_8\otimes M_{\textsl{y}}^{\textsl{b}})\rho_{\it{A_1A_2A_3B}}]$. The three sources send three classical messages $\lambda_i$ $(1\leq i\leq 3)$ to the central party $B$. One source generates a quantum state $\rho_{\lambda_3}^{\it{A_3}}$ ($\rho_{\lambda_1}^{\it{A_1}}$
or $\rho_{\lambda_2}^{\it{A_2}}$) with probability $p(\lambda_3)$ ($p(\lambda_1)$ or $p(\lambda_2)$). The other two sources generate entangled states $\rho_{\lambda_1\lambda_2}^{\it{A_1A_2}},$ $\rho_{\lambda_2\lambda_3}^{\it{A_2A_3}}$ or $\rho_{\lambda_1\lambda_3}^{\it{A_1A_3}}$ with the probabilities $p(\lambda_1,\lambda_2),$ $p(\lambda_2,\lambda_3)$ and $p(\lambda_1,\lambda_3)$ randomly, which are sent to the edge parties. The total probability that $\it{A_1A_2}$ ($\it{A_2A_3}$,  $\it{A_1A_3}$) receive the entangled local hidden states is $q_1$ ($q_2$, $q_3$).

\begin{figure}[H]
\includegraphics[width=9cm]{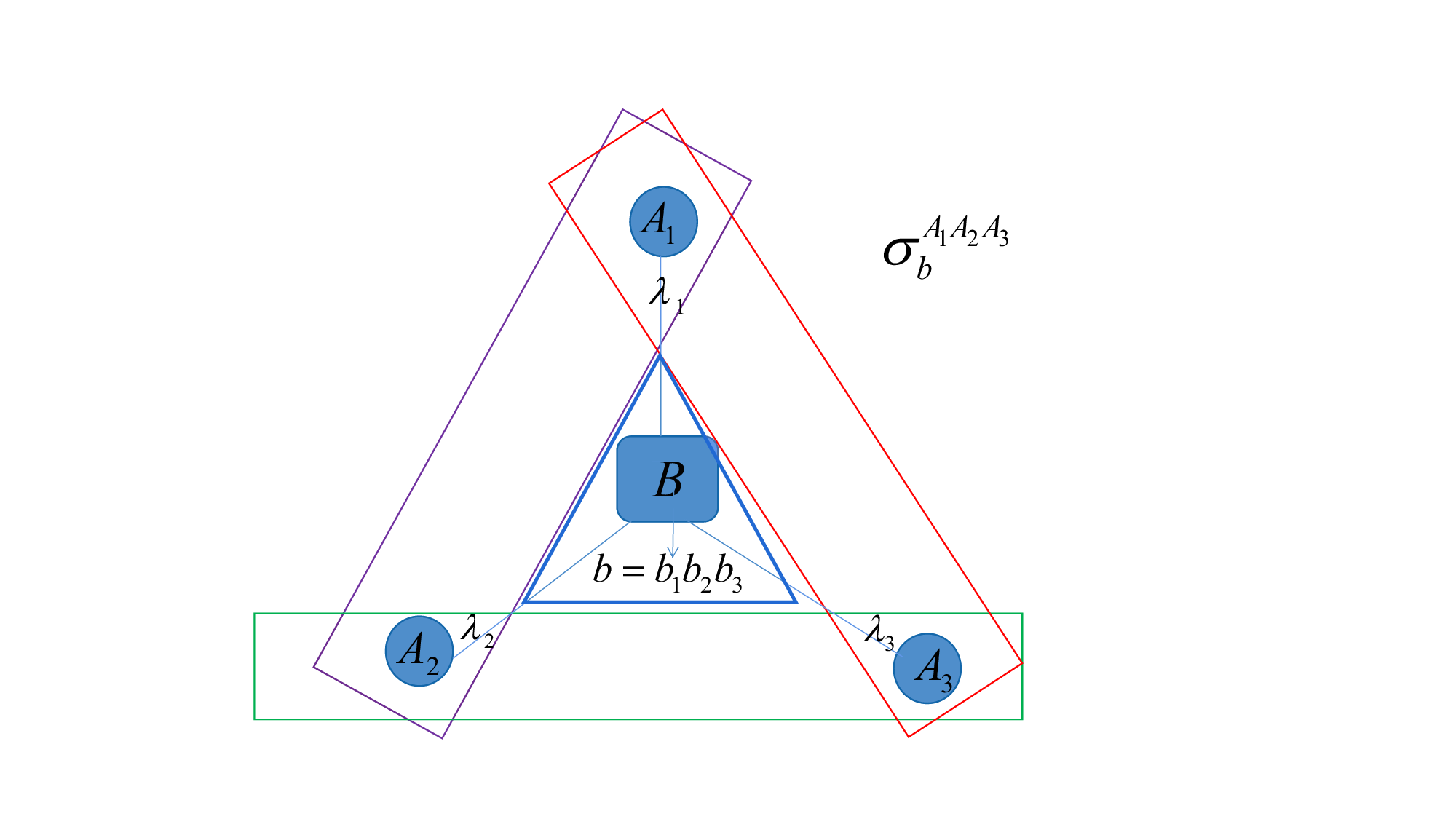}
\caption{The bi-separable local hidden state model of $\sigma_{\textsl{b}}^{\it{A_1A_2A_3}}$. $\sigma_{\textsl{b}}^{\it{A_1A_2A_3}}$ is represented by the region outside of the blue triangle. The entangled local hidden states $\rho_{\lambda_1\lambda_2}^{\it{A_1A_2}}$, $\rho_{\lambda_1\lambda_3}^{\it{A_1A_3}}$ and $\rho_{\lambda_2\lambda_3}^{\it{A_2A_3}}$ are shown in the purple, red and green boxes, respectively.}
\label{gen-steer}
\end{figure}

Then the sub-normalized state $\sigma_{\textsl{b}}^{\it{A_1A_2A_3}}$  admits bi-separable local hidden states (BLHS) if
\begin{eqnarray}
\begin{aligned}
&\sigma_{\textsl{b}}^{\it{A_1A_2A_3}}\\ \nonumber
=&q_1\sum\limits_{\lambda_1,\lambda_2,\lambda_3}p(\lambda_1,\lambda_2)
p(\lambda_3)p(\textsl{b}|\lambda_1,\lambda_2,\lambda_3)\rho_{\lambda_1\lambda_2}^{\it{A_1A_2}}
\otimes\rho_{\lambda_3}^{\it{A_3}}\\ \nonumber
+& q_2\sum\limits_{\lambda_1,\lambda_2,\lambda_3}p(\lambda_1)
p(\lambda_2,\lambda_3)p(\textsl{b}|\lambda_1,\lambda_2,\lambda_3)
\rho_{\lambda_2\lambda_3}^{\it{A_2A_3}}\otimes\rho_{\lambda_1}^{\it{A_1}}\\ \nonumber
+&q_3\sum\limits_{\lambda_1,\lambda_2,\lambda_3}p(\lambda_2)
p(\lambda_1,\lambda_3)p(\textsl{b}|\lambda_1,\lambda_2,\lambda_3)
\rho_{\lambda_1\lambda_3}^{\it{A_1A_3}}\otimes\rho_{\lambda_2}^{\it{A_2}}\\ \nonumber
\end{aligned}
\end{eqnarray}
with $\sum\limits_i q_i=1.$ The sketch of bi-separable local hidden states model of $\sigma_{\textsl{b}}^{\it{A_1A_2A_3}}$ is shown in Fig. (\ref{gen-steer}).
From the aspect of probabilities, we have that $\rho_{A_1A_2A_3B}$ admits network local hidden variable and bi-separable local hidden states (NLHV-BLHS) model if
\begin{eqnarray}
\begin{aligned}
&p(\textsl{a}_1,\textsl{a}_2, \textsl{a}_3,\textsl{b}|\textsl{x}_1,\textsl{x}_2,\textsl{x}_3)\\
=&q_1\sum\limits_{\lambda_1,\lambda_2,\lambda_3}p(\textsl{b}|\lambda_1,\lambda_2,\lambda_3)
p(\lambda_1,\lambda_2)p(\lambda_3)\\
&\times p_Q(\textsl{a}_1,\textsl{a}_2|\textsl{x}_1,\textsl{x}_2,\rho_{\lambda_1\lambda_2}^{\it{A_1A_2}})
p_Q(\textsl{a}_3|\textsl{x}_3,\rho_{\lambda_3}^{\it{A_3}})\\
+&q_2\sum\limits_{\lambda_1,\lambda_2,\lambda_3}p(\textsl{b}|\lambda_1,\lambda_2,\lambda_3)
p(\lambda_2,\lambda_3)p(\lambda_1)\\
&\times p_Q({\textsl{a}}_2,\textsl{a}_3|\textsl{x}_2,\textsl{x}_3,\rho_{\lambda_2\lambda_3}^{\it{A_2A_3}})
p_Q(\textsl{a}_1|\textsl{x}_1,\rho_{\lambda_1}^{\it{A_1}})\\
+&q_3\sum\limits_{\lambda_1,\lambda_2,\lambda_3}p(\textsl{b}|\lambda_1,\lambda_2,\lambda_3)
p(\lambda_1,\lambda_3)p(\lambda_2)\\
&\times p_Q(\textsl{a}_1,\textsl{a}_3|\textsl{x}_1,\textsl{x}_3,\rho_{\lambda_1\lambda_3}^{\it{A_1A_3}})
p_Q(\textsl{a}_2|\textsl{x}_2,\rho_{\lambda_2}^{\it{A_2}}),
\end{aligned}
\end{eqnarray}

Generally we have star networks with state $\rho_{A_1\cdots A_n B}=\bigotimes\limits_{i=1}^n \rho_{A_iB_i}$. Bob performs one fixed measurement $\textsl{y}$ defined in Theorem 2. It admits bi-separable local hidden states (BLHS) model if
\begin{eqnarray}
\begin{aligned}
&\sigma_{\textsl{b}}^{A_1\cdots A_n}\\
=&\sum\limits_{\lambda_1\cdots\lambda_n}p(\textsl{b}|\lambda_1,\cdots,\lambda_n)\\
&\times\sum\limits_{s=1}^{\lfloor\frac{n}{2}\rfloor} \sum\limits_{t=1}^{C_n^s}q^t_s p(\Lambda^t_s) \rho^{\Gamma^t_s}_{\Lambda^t_s}\otimes p(\tilde{\Lambda}^t_s)\rho^{\tilde{\Gamma}^t_s}_{\tilde{\Lambda}^t_s}
\end{aligned}
\end{eqnarray}
where
$\sum\limits_{s,t} q_s^t=1,$ and $\Lambda^t_s$ and $\tilde{\Lambda}^t_s$ are two disjoint subsets of the set $\{\lambda_1,\cdots,\lambda_n\}$ with the number of the elements $s$ and $n-s$ respectively.
The number of the sets $\Lambda_s^t$ is $C_n^s$ with $C_n^s$ being the combinations.
Let $\Lambda^t_s=\{\lambda_i|i\in K^t_s\}$ and $\tilde{\Lambda}^t_s=\{\lambda_i|i\in \overline{K^t_s}\}$ with $K^t_s$ being the set obtained by selecting $s$ elements from $\{1,2,\cdots,n\}$ and
$\overline{K^t_s}$ being the complement of $K^t_s.$
Then $\Gamma^t_s=\{\textit{A}_i|i\in K^t_s\}$ and $\tilde{\Gamma}^t_s=\{\textit{A}_i|i\in  \overline{K^t_s}\}.$

From the aspect of probabilities, we have $\rho_{A_1\cdots A_nB}$ admits network local hidden variable and bi-separable local hidden states (NLHV-BLHS) model if
\begin{eqnarray*}
\begin{aligned}
&p(\textsl{a}_1,\cdots,\textsl{a}_n,\textsl{b}|\textsl{x}_1,\cdots,\textsl{x}_n)\\
=&\sum\limits_{\lambda_1\cdots\lambda_n}p(\textsl{b}|\lambda_1,\cdots,\lambda_n)\sum\limits_{s=1}^{\lfloor\frac{n}{2}\rfloor}\sum\limits_{t=1}^{C_n^s} q_s^t p(\Lambda_s^t)p(\tilde{\Lambda}_s^t)\\
&\times p_Q(\textsl{a}_{K_s^t}|x_{K_s^t},\rho_{\Lambda_s^t}^{\Gamma_s^t})
p_Q(\textsl{a}_{\overline{K_s^t}}|x_{\overline{K_s^t}},\rho_{\tilde{\Lambda}_s^t}^{\tilde{\Gamma}_s^t}).
\end{aligned}
\end{eqnarray*}

The genuine multipartite entanglement of $\sigma_b^{\it{A_1\cdots A_n}}$ can rule out BLHS.
But it is also a difficult problem to detect the genuine entanglement.
In addition, BLHS model is different from the bi-separable state model as all the sub-normalized states in the network assemblage  admit the BLHS model with the same set of the bi-separable local hidden states. To investigate the BLHS model further, we give the witness to detect the genuine network steering composed of general quantum states, see proof in Appendix. Consider the state $\rho_{\it{A_1A_2A_3B}}$. The central party performs the fixed measurement $\Pi=\{G_{000},\cdots, G_{111}\}$ defined in Theorem 2. The edge parties $\it{A_i}$ perform the mutually unbiased observables $x_i^j.$ The state $\rho_{\it{A_1A_2A_3B}}$ admits BLHS if
\begin{eqnarray}
\begin{aligned}
&\sqrt{|\langle x_1^1\otimes x_2^1\otimes x_3^1\otimes y^{111}\rangle|}+\sqrt{|\langle x_1^1\otimes x_2^2\otimes x_3^2\otimes y^{122}\rangle|}\\ \nonumber
+&\sqrt{|\langle x_1^2\otimes x_2^1\otimes x_3^2\otimes y^{212}\rangle|}+\sqrt{|\langle x_1^2\otimes x_2^2\otimes x_3^1\otimes y^{221}\rangle|}\\ \nonumber
\leq& 2\sqrt{2}
\end{aligned}
\end{eqnarray}
and
\begin{eqnarray}
\begin{aligned}
&\sqrt{|\langle x_1^1\otimes x_2^1\otimes x_3^1\otimes y^{111}\rangle|}+\sqrt{|\langle x_1^1\otimes x_2^2\otimes x_3^2\otimes y^{122}\rangle|}\\ \nonumber
+&\sqrt{|\langle x_1^2\otimes x_2^1\otimes x_3^2\otimes y^{212}\rangle|}+\sqrt{|\langle x_1^2\otimes x_2^2\otimes x_3^1\otimes y^{221}\rangle|}\\ \nonumber
+&\sqrt{|\langle x_1^3\otimes x_2^3\otimes I_2\otimes y^{330}\rangle|}+\sqrt{|\langle x_1^3\otimes I_2\otimes x_3^3\otimes y^{303}\rangle|}\\ \nonumber
+&\sqrt{|\langle I_2\otimes x_2^3\otimes x_3^3\otimes y^{033}\rangle|}\leq 2\sqrt{3}+1.
\end{aligned}
\end{eqnarray}

For the star network $\rho_{\it{A_1\cdots A_n B}}$, we have the following result, see proof in Appendix.

{\bf{Theorem 4.}} If  $\rho_{\it{A_1\cdots A_n B}}$ admits NLHV-BLHS model from the central party to the edge parties when Bob performs the fixed measurement given in Theorem 2, we have
\begin{eqnarray}
\begin{aligned}
\label{2-set-gen}
&\sum\limits_{i_1,\cdots,i_n\in C}|\langle x^{i_1}_1\otimes x^{i_2}_2\otimes \cdots \otimes x^{i_n}_n\otimes y^{i_1\cdots i_n}\rangle|^{\frac{1}{2}}\\
&\leq 2^{n-2}\sqrt{2}
\end{aligned}
\end{eqnarray}
and
\begin{eqnarray}
\begin{aligned}
\label{3-set-gen}
&\sum\limits_{i_1,\cdots,i_n\in C\cup C'}|\langle x^{i_1}_1\otimes x^{i_2}_2\otimes \cdots \otimes x^{i_n}_n\otimes y^{i_1\cdots i_n}\rangle|^{\frac{1}{2}}\\
&\leq 2^{n-2}\sqrt{3}+2^{n-2}-1.
\end{aligned}
\end{eqnarray}

As an example,  let us consider the star network $\rho_{\it{A_1\cdots A_n B}}=\rho_{\it{A_1B_1}}\otimes \cdots\otimes\rho_{\it{A_nB_n}}$, where $\rho_{\it{A_iB_i}}=\frac{1}{4}(\rm{I}_4+c_1\sigma_1\otimes\sigma_1
+c_2\sigma_2\otimes\sigma_2+c_3\sigma_3\otimes\sigma_3)$ and $\it{B}=\it{B_1B_2\cdots B_n}$.
We have that $\rho_{\it{A_1\cdots A_n B}}$ is genuine steerable from the central party to the edge parties when
$$\sum\limits_{i_1\cdots i_n\in C}\sqrt[2]{|c_{i_1}\cdots c_{i_n}|}>2^{n-2}\sqrt{2}$$
in two measurement settings by using the inequality (\ref{2-set-gen}),
and $$\sum\limits_{i_1\cdots i_n\in C\bigcup C'}\sqrt[2]{|c_{i_1}\cdots c_{i_n}|}>2^{n-2}\sqrt{3}+2^{n-2}-1$$
in three measurement settings with $c_0=1$ by using the inequality (\ref{3-set-gen}).

Specially, when $|c_1|=|c_2|=|c_3|=p$, $\rho_{\it{A_1\cdots A_n B}}$ is genuine steerable for $p>\frac{1}{\sqrt[n]{2}}$ in two measurement settings and for $p>p_0$ in three measurement settings, where $p_0$ the solutions of $2^{n-1}p^{\frac{n}{2}}+C_n^{2}p^{\frac{2}{2}}+\cdots
+C_n^{n}p^{\frac{n}{2}}=2^{n-2}\sqrt{3}+2^{n-2}-1$ ($n$ is even) or $2^{n-1}p^{\frac{n}{2}}+C_n^{2}p^{\frac{2}{2}}+\cdots
+C_n^{n-1}p^{\frac{n-1}{2}}=2^{n-2}\sqrt{3}+2^{n-2}-1$ ($n$ is odd),
$p_0=0.703$ when $n=3$ and $p_0=0.769$ when $n=4$.
However, since $\sigma_b^{\it{A_1A_2A_3}}/\rm{Tr}[\sigma_b^{\it{A_1A_2A_3}}]$ is genuine tripartite entangled if and only if $p>0.7154$, interestingly we have that the separable assemblage may have genuine network steering.

If $\rho_{A_iB_i}=\frac{1}{4}(\rm{I}+\sum\limits_j (a_j\sigma_j\otimes \rm{I}_2+ b_j\rm{I}_2\otimes \sigma_j)+\sum\limits_{kl}c_{kl}\sigma_k\otimes \sigma_l)$,
which is locally unitary equivalent to $\rho_{\it{A_iB_i}}=\frac{1}{4}(\rm{I}+\sum\limits_j (a'_j\sigma'_j\otimes \rm{I}_2+ b'_j\rm{I}_2\otimes \sigma'_j)+\sum\limits_{k}d_{k}\sigma'_k\otimes \sigma'_k)$, we have that $\rho_{\it{A_1\cdots A_nB}}$ is genuine steerable from the central party to the edge parties when $$\sum\limits_{i_1\cdots i_n\in C}\sqrt[2]{|d_{i_1}\cdots d_{i_n}|}>2^{n-2}\sqrt{2}$$ in two measurement settings,
and $$\sum\limits_{i_1\cdots i_n\in C\bigcup C'}\sqrt[2]{|d_{i_1}\cdots d_{i_n}|}>2^{n-2}\sqrt{3}+2^{n-2}-1$$
in three measurement settings.

The genuine network steering can be detected in the star network scenarios composed of bell-diagonal states and general two-qubit states by the inequalities we derived in Theorem 4. Interestingly, the bi-separable states under the joint measurements performed by the central party do not admit BLHS model for the star network composed of Werner states. 

\section{Conclusion}
We have investigated the network steering and genuine network steering in star network scenarios from the aspect of the probabilities, when the central party performs the fixed measurement. We have constructed the linear and nonlinear inequalities to verify the network steering and the genuine network steering in two and three measurement settings. It has been shown that more quantum network steering can be detected compared with the violation of the $n$-locality quantum networks, and the biseparable assemblages may show genuine network steering in star network configurations. It would be also interesting to explore the applications of the network genuine steering in quantum processing tasks. Our results may also highlight further investigations on multipartite quantum network steering and genuine multipartite quantum network steering in high dimensional systems.

\bigskip

\noindent{\bf ACKNOWLEDGEMENTS}\, \, This work is supported by the National Natural Science Foundation of China (NSFC) under Grants  12071179, 12075159 and 12171044; the Academician Innovation Platform of Hainan Province.

\section{Appendix}
{\bf{Theorem 1.}} For the simple network, the probabilities that admit the NLHS-LHV model from Bob to Alice and Charlie satisfy the following inequalities,
\begin{eqnarray}\label{s-1p}
\sum_{i=1}^3 |\langle x^i\otimes y^i\otimes z^i\rangle|\leq 1.
\end{eqnarray}

\noindent{\bf{Proof of Theorem 1:}}

By definition if Bob can not steer Alice and Charlie, we have the following inequality,
\begin{eqnarray}
\begin{aligned}
&|\langle x^1\otimes y^1 \otimes z^1\rangle|+|\langle x^2\otimes y^2 \otimes z^2\rangle|+|\langle x^3\otimes y^3 \otimes z^3\rangle|\\ \nonumber
=|&\sum\limits_{\lambda_1,\lambda_2}p(\lambda_1)p(\lambda_2)\sum\limits_{a}(-1)^a p_Q(a|\lambda_1,x^1)\\ \nonumber
&\times\sum\limits_{b}(-1)^{b_2}p(b|\lambda_1,\lambda_2)\sum\limits_{c}(-1)^c p_Q(c|\lambda_2,z^1)|\\ \nonumber
&+|\sum\limits_{\lambda_1,\lambda_2}p(\lambda_1)p(\lambda_2)\sum\limits_{a}(-1)^a p_Q(a|\lambda_1,x^2)\\ \nonumber
&\times\sum\limits_{b}(-1)^{b_1}p(b|\lambda_1,\lambda_2)\sum\limits_{c}(-1)^c p_Q(c|\lambda_2,z^2)|\\ \nonumber
&+|\sum\limits_{\lambda_1,\lambda_2}p(\lambda_1)p(\lambda_2)\sum\limits_{a}(-1)^a p_Q(a|\lambda_1,x^3)\\ \nonumber
&\times\sum\limits_{b}(-1)^{b_1+b_2} p(b|\lambda_1,\lambda_2,y_2)\sum\limits_{c}(-1)^c p_Q(c|\lambda_2,z^3)|\\ \nonumber
=&|\sum\limits_{\lambda_1,\lambda_2}p(\lambda_1)p(\lambda_2)\langle x^1\rangle^Q_{\lambda_1}\langle y^1\rangle_{\lambda_1,\lambda_2}\langle z^1\rangle^Q_{\lambda_2}|\\ \nonumber
+&|\sum\limits_{\lambda_1,\lambda_2}p(\lambda_1)p(\lambda_2)\langle x^2\rangle^Q_{\lambda_1}\langle y^2\rangle_{\lambda_1,\lambda_2}\langle z^2\rangle^Q_{\lambda_2}|\\ \nonumber
+&|\sum\limits_{\lambda_1,\lambda_2}p(\lambda_1)p(\lambda_2)\langle x^3\rangle^Q_{\lambda_1}\langle y^3\rangle_{\lambda_1,\lambda_2}\langle z^3\rangle^Q_{\lambda_2}|\\ \nonumber
\leq & \sum\limits_{\lambda_1,\lambda_2}p(\lambda_1)p(\lambda_2)|\langle x^1\rangle^Q_{\lambda_1}\langle y^1\rangle_{\lambda_1,\lambda_2}\langle z^1\rangle^Q_{\lambda_2}|\\ \nonumber
+& \sum\limits_{\lambda_1,\lambda_2}p(\lambda_1)p(\lambda_2)|\langle x^2\rangle^Q_{\lambda_1}\langle y^2\rangle_{\lambda_1,\lambda_2}\langle z^2\rangle^Q_{\lambda_2}|\\ \nonumber
+& \sum\limits_{\lambda_1,\lambda_2}p(\lambda_1)p(\lambda_2)|\langle x^3\rangle^Q_{\lambda_1}\langle y^3\rangle_{\lambda_1,\lambda_2}\langle z^3\rangle^Q_{\lambda_2}|\\ \nonumber
\leq &\sum\limits_{\lambda_1,\lambda_2}p(\lambda_1)p(\lambda_2)\sqrt{(\langle x^1\rangle^Q_{\lambda_1})^2+(|\langle x^2\rangle^Q_{\lambda_1})^2+(|\langle x^3\rangle^Q_{\lambda_1})^2}\\ \nonumber
&\times \sqrt{(\langle z^1\rangle^Q_{\lambda_2})^2+(|\langle z^2\rangle^Q_{\lambda_2})^2+(|\langle z^3\rangle^Q_{\lambda_2})^2}\leq 1,
\end{aligned}
\end{eqnarray}
where the equality is due to the definitions of the expectation and the network local hidden variable and local hidden state model. The first inequality is due to the inequality satisfied by absolute values, namely, $|\sum\limits_i s_i|\leq \sum\limits_i |s_i|$. The second inequality is due to the inequality $|\langle y^i\rangle_{\lambda_1,\lambda_2}|\leq 1$ and the Cauchy-Schwarz inequality. The last inequality is from that $(\langle x^1\rangle^Q_{\lambda_1,\lambda_2})^2+(\langle x^2\rangle^Q_{\lambda_1,\lambda_2})^1+(\langle x^3\rangle^Q_{\lambda_1,\lambda_2})^2
\leq 1$ and $(\langle z^1\rangle_{\lambda_2}^Q)^2+(\langle z^2\rangle_{\lambda_2}^Q)^2+(\langle z^3\rangle_{\lambda_2}^Q)^2
\leq 1$ for the mutually unbiased measurements.

\bigskip

\noindent{\bf{Theorem 2:}} {\bf{(a)}} If $\rho_{AB}=\bigotimes\limits_{k=1}^n\rho_{A_{k}B_{k}}$
admits NLHV-LHS from the central parties to the edge parties when Bob performs the fixed measurement in the two measurement settings, we have\\
(1) When $n$ is an odd number,
\begin{eqnarray}
\begin{aligned}
\label{2-set-nodd-1p}
&\sum\limits_{i_1,\cdots,i_n\in C}|\langle x^{i_1}_1\otimes x^{i_2}_2\otimes \cdots \otimes x^{i_n}_n\otimes y^{i_1\cdots i_n}\rangle|^{\frac{2}{n}}\\
&\leq 2^{n-2}
\end{aligned}
\end{eqnarray}
and
\begin{eqnarray}
\begin{aligned}
\label{2-set-nodd-2p}
&\sum\limits_{i_1,\cdots,i_n\in C}|\langle x^{i_1}_1\otimes x^{i_2}_2\otimes \cdots \otimes x^{i_n}_n\otimes y^{i_1\cdots i_n}\rangle|^{\frac{1}{n}}\\
&\leq 2^{n-2}\sqrt{2},
\end{aligned}
\end{eqnarray}

\noindent{\bf{Proof of Theorem 2}} {(\bf{a})} (1): If $\rho_{AB}=\bigotimes\limits_{k=1}^n\rho_{A_{k}B_{k}}$
admits NLHV-LHS from the central parties to the edge parties when Bob performs the fixed measurement in the two measurement settings, then
\begin{eqnarray}
\begin{aligned}
&\sum\limits_{i_1,\cdots,i_n\in C}|\langle x^{i_1}_1\otimes x^{i_2}_2\otimes \cdots \otimes x^{i_n}_n\otimes y^{i_1\cdots i_n}\rangle|^{\frac{2}{n}}\\ \nonumber
&= \sum\limits_{i_1,\cdots,i_n\in C}|\sum\limits_{\lambda_k}\sum\limits_{a_k^{i_k}}\prod\limits_{k=1}^n p(\lambda_k)\sum\limits_{a_k^{i_k}}(-1)^{a_k^{i_k}} p_Q(a_k^{i_k}|x_k^{i_k},\lambda_k)\\ \nonumber
&\times \sum\limits_b (-1)^{\Re} p(b|\lambda_1\cdots\lambda_n)|^{\frac{2}{n}}\\ \nonumber
&=\sum\limits_{i_1,\cdots,i_n\in C}|\sum\limits_{\lambda_k}\prod\limits_{k=1}^n p(\lambda_k)\langle x_k^{i_k}\rangle_{\lambda_k}^Q\langle y^{i_1\cdots i_n}\rangle_{\lambda_1\cdots\lambda_n}|)^{\frac{2}{n}} \\ \nonumber
&\leq \sum\limits_{i_1,\cdots,i_n\in C}\sum\limits_{\lambda_k}(\prod\limits_{k=1}^n p(\lambda_k)|\langle x_k^{i_k}\rangle_{\lambda_k}^Q|)^{\frac{2}{n}}\\ \nonumber
&\leq \prod\limits_{k=1}^n \{\sum\limits_{\lambda_k}p(\lambda_k)\sum\limits_{i_k\in C}[\langle x_k^{i_k}\rangle_{\lambda_k}^Q]^2\}^{\frac{1}{n}}\\ \nonumber
&\leq 2^{n-2},
\end{aligned}
\end{eqnarray}
where the equalities are attained according to the definitions of expectation value and NLHV-LHS model, the first inequality is attained
by using the absolute value inequality and $|\langle y^{i_1\cdots i_n}\rangle_{\lambda_1\cdots\lambda_n}|\leq 1,$
the second inequality is attained by using the Lemma 1 in \cite{Tavakoli}.
There are $2^{n-1}$ elements in $C$. Hence, the number of $1$ and the number of $2$ are all $2^{n-2}$ for each $k$. Then we have $\sum\limits_{i_k\in C}[\langle x_k^{i_k}\rangle_{\lambda_k}^Q]^2\leq 2^{n-2}$ by using $({\langle x_k^1\rangle^Q_{\lambda_k}})^2+({\langle x_k^2\rangle^Q_{\lambda_k}})^2\leq 1$ for the mutually unbiased measurements, which gives rise to the last inequality.

{\bf Theorem 2:} {(\bf{a})} (2) When $n$ is an even number,
\begin{eqnarray}
\begin{aligned}
\label{2-set-neven-1p}
&|\langle x^{1}_1\otimes x^{1}_2\otimes \cdots \otimes x^{1}_n\otimes y_1^1\rangle|^{\frac{2}{n}}\\
+&|\langle x^{2}_1\otimes x^{2}_2\otimes \cdots \otimes x^{2}_n\otimes y_1^2\rangle|^{\frac{2}{n}}\leq 1.
\end{aligned}
\end{eqnarray}
and
\begin{eqnarray}
\begin{aligned}
\label{2-set-neven-2p}
&|\langle x^{1}_1\otimes x^{1}_2\otimes \cdots \otimes x^{1}_n\otimes y_1^1\rangle|^{\frac{1}{n}}\\
+&|\langle x^{2}_1\otimes x^{2}_2\otimes \cdots \otimes x^{2}_n\otimes y_1^2\rangle|^{\frac{1}{n}}\leq \sqrt{2}.
\end{aligned}
\end{eqnarray}

{\bf{Proof of Theorem 2}}. {(\bf{a})} (2): By straightforward calculation,
\begin{eqnarray}
\begin{aligned}
&|\langle\bigotimes\limits_{k=1}^n x_k^1\otimes y^1_1\rangle|^{\frac{2}{n}}+|\langle\bigotimes\limits_{k=1}^n x_k^2\otimes y_1^2\rangle|^{\frac{2}{n}}\\ \nonumber
=&(|\sum\limits_{\lambda_k}\prod\limits_{k=1}^n p(\lambda_k)\sum\limits_{a_k^1}(-1)^{a_k^1} p_Q(a_k^1|x_k^1,\lambda_k) \sum\limits_b(-1)^{\Re}\\ \nonumber
 &\times p(b|\lambda_1\cdots\lambda_n)|)^{\frac{2}{n}}\\ \nonumber
+&(|\sum\limits_{\lambda_k}\prod\limits_{k=1}^n p(\lambda_k)\sum\limits_{a_k^2}(-1)^{a_k^2} p_Q(a_k^2|x_k^2,\lambda_k) \sum\limits_b(-1)^{\Re} \\ \nonumber
&\times p(b|\lambda_1\cdots\lambda_n)|)^{\frac{2}{n}}\\ \nonumber
=&(|\sum\limits_{\lambda_k}\prod\limits_{k=1}^n p(\lambda_k)\langle x_k^1\rangle_{\lambda_k}^Q\langle y^1_1\rangle_{\lambda}|)^{\frac{2}{n}}\\
+&(|\sum\limits_{\lambda_k}\prod\limits_{k=1}^n p(\lambda_k)\langle x_k^2\rangle^Q_{\lambda_k}\langle y_1^2\rangle_{\lambda}|)^{\frac{2}{n}}\\ \nonumber
\leq &\sum\limits_{\lambda_k}(|\prod\limits_{k=1}^n p(\lambda_k) |\langle x_k^1\rangle_{\lambda_k}^Q\langle y_1^1\rangle_{\lambda}|)^{\frac{2}{n}}\\
&+\sum\limits_{\lambda_k}(|\prod\limits_{k=1}^n p(\lambda_k) |\langle x_k^2\rangle_{\lambda_k}^Q\langle y_1^2\rangle_{\lambda}|)^{\frac{2}{n}}\\ \nonumber
\leq & \prod\limits_{k=1}^n(\sum\limits_{\lambda_k}  p(\lambda_k) ((\langle x_k^1\rangle^Q_{\lambda_k})^2+(\langle x_k^2\rangle^Q_{\lambda_k})^2)^{\frac{1}{n}}\\ \nonumber
\leq & 1,
\end{aligned}
\end{eqnarray}
where the equalities are attained by using the definition of expectation values and NLHV-LHS model, the second inequality is attained by using the Lemma 1 in \cite{Tavakoli}, the last inequality uses $({\langle x_k^1\rangle^Q_{\lambda_k}})^2+({\langle x_k^2\rangle^Q_{\lambda_k}})^2\leq 1$ for the mutually unbiased measurements.

{\bf{Theorem 2}}{\bf{(b)}} If $\rho_{AB}=\bigotimes\limits_{k=1}^n\rho_{A_{k}B_{k}}$ admits NLHV-LHS from the central parties to the edge parties when Bob performs the fixed measurement in three measurement settings, we have\\
(1) When $n$ is an odd number,
\begin{eqnarray}
\begin{aligned}
\label{3-set-nodd-1p}
&\sum\limits_{i_1,\cdots,i_n\in C\cup C'}|\langle x^{i_1}_1\otimes x^{i_2}_2\otimes \cdots \otimes x^{i_n}_n\otimes y^{i_1\cdots i_n}\rangle|^{\frac{2}{n}}\\
&\leq 2^{n-2}+2^{n-2}-1
\end{aligned}
\end{eqnarray}
and
\begin{eqnarray}
\begin{aligned}
\label{3-set-nodd-2p}
&\sum\limits_{i_1,\cdots,i_n\in C\cup C'}|\langle x^{i_1}_1\otimes x^{i_2}_2\otimes \cdots \otimes x^{i_n}_n\otimes y^{i_1\cdots i_n}\rangle|^{\frac{1}{n}}\\
&\leq 2^{n-2}\sqrt{3}+2^{n-2}-1.
\end{aligned}
\end{eqnarray}

{\bf{Proof of Theorem 2}}{(\bf{b})} (1).
\begin{eqnarray}
\begin{aligned}
&\sum\limits_{i_1,\cdots,i_n\in C\bigcup C'}|\langle x^{i_1}_1\otimes x^{i_2}_2\otimes \cdots \otimes x^{i_n}_n\otimes y_1^{i_1\cdots i_n}\rangle|^{\frac{2}{n}}\\ \nonumber
&\leq \sum\limits_{i_1,\cdots,i_n\in C\bigcup C'}|\sum\limits_{\lambda_k}\prod\limits_{k=1}^n p(\lambda_k)(-1)^{a_k^{i_k}} p_Q(a_k^{i_k}|x_k^{i_k},\lambda_k)\\ \nonumber
&\times (-1)^{\Re} p(b|\lambda_1\cdots\lambda_n)|^{\frac{2}{n}}\\ \nonumber
&=\sum\limits_{i_1,\cdots,i_n\in C\bigcup C'}|\sum\limits_{\lambda_k}\prod\limits_{k=1}^n p(\lambda_k)\langle x_k^{i_k}\rangle_{\lambda_k}^Q|)^{\frac{2}{n}} \\ \nonumber
&\leq \sum\limits_{i_1,\cdots,i_n\in C\bigcup C'}\sum\limits_{\lambda_k}(\prod\limits_{k=1}^n p(\lambda_k)|\langle x_k^{i_k}\rangle_{\lambda_k}^Q|)^{\frac{2}{n}}\\ \nonumber
&\leq \prod\limits_{k=1}^n \{\sum\limits_{\lambda_k}p(\lambda_k)\sum\limits_{i_k\in C\bigcup C'}[\langle x_k^{i_k}\rangle_{\lambda_k}^Q]^2\}^{\frac{1}{n}}\\ \nonumber
&\leq 2^{n-2}+C_{n-1}^2+\cdots+C_{n-1}^{n-1}=2^{n-2}+2^{n-2}-1.
\end{aligned}
\end{eqnarray}
 The proof is similar to that of ${(\bf{a})}(1),$ but the last inequality is attained due to that the number of $3$ in $C'$ is $2^{n-2}$ and the number of $0$ in $C'$
is $C_{n-1}^2+\cdots+C_{n-1}^{n-1}$ for each $k$.

{\bf{Theorem 2}}{(\bf{b})}(2) When $n$ is an even number,
\begin{eqnarray}
\begin{aligned}
\label{3-set-neven-1p}
&|\langle x^{1}_1\otimes x^{1}_2\otimes \cdots \otimes x^{1}_n\otimes y_1^1\rangle|^{\frac{2}{n}}\\
+&|\langle x^{2}_1\otimes x^{2}_2\otimes \cdots \otimes x^{2}_n\otimes y_1^2\rangle|^{\frac{2}{n}}\\
+&|\langle x^{3}_1\otimes x^{3}_2\otimes \cdots \otimes x^{3}_n\otimes y_1^3\rangle|^{\frac{2}{n}}\leq 1
\end{aligned}
\end{eqnarray}
and
\begin{eqnarray}
\begin{aligned}
\label{3-set-neven-2p}
&|\langle x^{1}_1\otimes x^{1}_2\otimes \cdots \otimes x^{1}_n\otimes y_1^1\rangle|^{\frac{1}{n}}\\
+&|\langle x^{2}_1\otimes x^{2}_2\otimes \cdots \otimes x^{2}_n\otimes y_1^2\rangle|^{\frac{1}{n}}\\
+&|\langle x^{3}_1\otimes x^{3}_2\otimes \cdots \otimes x^{3}_n\otimes y_1^3\rangle|^{\frac{1}{n}}\leq \sqrt{3},
\end{aligned}
\end{eqnarray}
where $x_k^{i_k}$ $(i_k=1,2,3)$ are all mutually unbiased measurements for $k=1,\cdots,n,$
$x_1^0=x_2^0=\cdots x_n^0=\rm{I}_2$ are the identity matrices.

{\bf{Proof of Theorem 2}} {\bf{b}}(2). By straightforward calculation, we have
\begin{eqnarray}
\begin{aligned}
&|\langle\bigotimes\limits_{k=1}^n x_k^1\otimes y^1_1\rangle|^{\frac{2}{n}}+|\langle\bigotimes\limits_{k=1}^n x_k^2\otimes y_1^2\rangle|^{\frac{2}{n}}+|\langle\bigotimes\limits_{k=1}^n x_k^3\otimes y_1^3\rangle|^{\frac{2}{n}}\\ \nonumber
=&(|\sum\limits_{\lambda_k}\prod\limits_{k=1}^n p(\lambda_k)(-1)^{a_k^1} p_Q(a_k^1|x_k^1,\lambda_k) (-1)^{\Re}\\ \nonumber
 &\times p(b|\lambda_1\cdots\lambda_n)|)^{\frac{2}{n}}\\ \nonumber
+&(|\sum\limits_{\lambda_k}\prod\limits_{k=1}^n p(\lambda_k)(-1)^{a_k^2} p_Q(a_k^2|x_k^2,\lambda_k) (-1)^{\Re} \\ \nonumber
&\times p(b|\lambda_1\cdots\lambda_n)|)^{\frac{2}{n}}\\ \nonumber
+&(|\sum\limits_{\lambda_k}\prod\limits_{k=1}^n p(\lambda_k)(-1)^{a_k^3} p_Q(a_k^3|x_k^3,\lambda_k) (-1)^{\Re} \\ \nonumber
&\times p(b|\lambda_1\cdots\lambda_n)|)^{\frac{2}{n}}\\ \nonumber
=&(|\sum\limits_{\lambda_k}\prod\limits_{k=1}^n p(\lambda_k)\langle x_k^1\rangle_{\lambda_k}^Q\langle y_1^1\rangle_{\lambda}|)^{\frac{2}{n}}\\
+&(|\sum\limits_{\lambda_k}\prod\limits_{k=1}^n p(\lambda_k)\langle x_k^2\rangle^Q_{\lambda_k}\langle y_1^2\rangle_{\lambda}|)^{\frac{2}{n}}\\ \nonumber
+&(|\sum\limits_{\lambda_k}\prod\limits_{k=1}^n p(\lambda_k)\langle x_k^3\rangle^Q_{\lambda_k}\langle y_1^3\rangle_{\lambda}|)^{\frac{2}{n}}\\ \nonumber
\leq &\sum\limits_{\lambda_k}(|\prod\limits_{k=1}^n p(\lambda_k) |\langle x_k^1\rangle_{\lambda_k}^Q\langle y^1_1\rangle_{\lambda}|)^{\frac{2}{n}}\\
&+\sum\limits_{\lambda_k}(|\prod\limits_{k=1}^n p(\lambda_k) |\langle x_k^2\rangle_{\lambda_k}^Q\langle y_1^2\rangle_{\lambda}|)^{\frac{2}{n}}\\ \nonumber
&+\sum\limits_{\lambda_k}(|\prod\limits_{k=1}^n p(\lambda_k) |\langle x_k^3\rangle_{\lambda_k}^Q\langle y_1^3\rangle_{\lambda}|)^{\frac{2}{n}}\\ \nonumber
\leq & \prod\limits_{k=1}^n(\sum\limits_{\lambda_k}  p(\lambda_k) ((\langle x_k^1\rangle^Q_{\lambda_k})^2+(\langle x_k^2\rangle^Q_{\lambda_k})^2+(\langle x_k^3\rangle^Q_{\lambda_k})^2)^{\frac{1}{n}}\\ \nonumber
\leq & 1,
\end{aligned}
\end{eqnarray}
where the equalities are attained by using the definition of expectation values and the NLHV-LHS model. The first inequality is attained by using the absolute value inequality. The second inequality is attained by using the Lemma 1 in \cite{Tavakoli}, the last inequality uses $({\langle x_k^1\rangle^Q_{\lambda_k}})^2+({\langle x_k^2\rangle^Q_{\lambda_k}})^2+({\langle x_k^3\rangle^Q_{\lambda_k}})^2\leq 1.$
The proof of inequalities (\ref{2-set-nodd-2}),(\ref{2-set-neven-2}),(\ref{3-set-nodd-2}) and (\ref{3-set-neven-2}) is similar,
but we use $({\langle x_k^1\rangle^Q_{\lambda_k}})+({\langle x_k^2\rangle^Q_{\lambda_k}})\leq \sqrt{2}$ and $({\langle x_k^1\rangle^Q_{\lambda_k}})+({\langle x_k^2\rangle^Q_{\lambda_k}})+({\langle x_k^3\rangle^Q_{\lambda_k}})\leq \sqrt{3}$ in the last step.

\bigskip
{\bf{Theorem 3.}} The star network state $\rho_{A_1\cdots A_n B}$ admits NLHS model from the central party to the edge parties if
\begin{eqnarray}
\begin{aligned}
|J_1|^{\frac{1}{n}}+|J_2|^{\frac{1}{n}}+|J_3|^{\frac{1}{n}}+|J_4|^{\frac{1}{n}}\leq 4,
\end{aligned}
\end{eqnarray}
where
\begin{eqnarray}
\begin{aligned}
&J_1=\langle\bigotimes\limits_{i=1}^n(x_i^1+x_i^2+x_i^3)\otimes B_1\rangle, \nonumber\\
&J_2=\langle\bigotimes\limits_{i=1}^n(x_i^1+x_i^2-x_i^3)\otimes B_2\rangle,\nonumber\\
&J_3=\langle\bigotimes\limits_{i=1}^n(x_i^1-x_i^2+x_i^3)\otimes B_3\rangle, \nonumber\\
&J_4=\langle\bigotimes\limits_{i=1}^n(-x_i^1+x_i^2+x_i^3)\otimes B_4\rangle \nonumber\\
\end{aligned}
\end{eqnarray}
and $x_i^{j}$ $(i=1,\cdots, n,\,j=1,2,3)$ are all mutually unbiased measurements.

\noindent{\bf{Proof of Theorem 3:}}

If $\rho_{AB}=\bigotimes\limits_{k=1}^n\rho_{A_kB_k}$ admits NLHV-LHS model from the central parties to the edge parties in three measurement
settings, using the definite of NLHV-NLHS model and $|\langle B_i\rangle|\leq 1$, we have
\begin{eqnarray}
\begin{aligned}
&|J_1|=\bigotimes\limits_{i=1}^n|(\langle x_i^1\rangle+\langle x_i^2\rangle+\langle x_i^3\rangle)\langle B_1\rangle|\\ \nonumber
&\quad\quad\leq \bigotimes\limits_{i=1}^n|\langle x_i^1\rangle+\langle x_i^2\rangle+\langle x_i^3\rangle|,\\ \nonumber
&|J_2|=\bigotimes\limits_{i=1}^n|(\langle x_i^1\rangle+\langle x_i^2\rangle-\langle x_i^3\rangle)\langle B_2\rangle|\\ \nonumber
&\quad\quad\leq \bigotimes\limits_{i=1}^n|\langle x_i^1\rangle+\langle x_i^2\rangle-\langle x_i^3\rangle|,\\ \nonumber
&|J_3|=\bigotimes\limits_{i=1}^n|(\langle x_i^1\rangle-\langle x_i^2\rangle+\langle x_i^3\rangle)\langle B_3\rangle|\\ \nonumber
&\quad\quad\leq \bigotimes\limits_{i=1}^n|\langle x_i^1\rangle-\langle x_i^2\rangle+\langle x_i^3\rangle|,\\ \nonumber
&|J_4|=\bigotimes\limits_{i=1}^n|(-\langle x_i^1\rangle+\langle x_i^2\rangle+\langle x_i^3\rangle)\langle B_4\rangle|\\ \nonumber
&\quad\quad\leq \bigotimes\limits_{i=1}^n|-\langle x_i^1\rangle+\langle x_i^2\rangle+\langle x_i^3\rangle|\nonumber
\end{aligned}
\end{eqnarray}
and hence
\begin{eqnarray}
\begin{aligned}
&|J_1|^{\frac{1}{n}}+|J_2|^{\frac{1}{n}}+|J_3|^{\frac{1}{n}}+|J_4|^{\frac{1}{n}}\\
\leq &\bigotimes\limits_{i=1}^n[|\langle x_i^1\rangle+\langle x_i^2\rangle+\langle x_i^3\rangle|+|\langle x_i^1\rangle+\langle x_i^2\rangle-\langle x_i^3\rangle|\\ \nonumber
+&|\langle x_i^1\rangle-\langle x_i^2\rangle+\langle x_i^3\rangle|+|-\langle x_i^1\rangle+\langle x_i^2\rangle+\langle x_i^3\rangle|]^{\frac{1}{n}}\\ \nonumber
\leq &4,
\end{aligned}
\end{eqnarray}
where the first inequality is attained by using Lemma 1 in \cite{Tavakoli}, and the second inequality is attained by using $|\langle x_i^1\rangle\pm\langle x_i^2\rangle\pm\langle x_i^3\rangle|=\pm(\langle x_i^1\rangle\pm\langle x_i^2\rangle\pm\langle x_i^3\rangle).$ 
\bigskip

{\bf{Theorem 4.}} If  $\rho_{\it{A_1\cdots A_n B}}$ admits NLHV-BLHS model from the central party to the edge parties when Bob performs the fixed measurement given in Theorem 2 and the edge parties perform mutually unbiased measurements, we have
\begin{eqnarray}
\begin{aligned}
\label{2-set-genp}
&\sum\limits_{i_1,\cdots,i_n\in C}|\langle x^{i_1}_1\otimes x^{i_2}_2\otimes \cdots \otimes x^{i_n}_n\otimes y_1^{i_1\cdots i_n}\rangle|^{\frac{1}{2}}\\
&\leq 2^{n-2}\sqrt{2}
\end{aligned}
\end{eqnarray}
and
\begin{eqnarray}
\begin{aligned}
\label{3-set-genp}
&\sum\limits_{i_1,\cdots,i_n\in C\cup C'}|\langle x^{i_1}_1\otimes x^{i_2}_2\otimes \cdots \otimes x^{i_n}_n\otimes y_1^{i_1\cdots i_n}\rangle|^{\frac{1}{2}}\\
&\leq 2^{n-2}\sqrt{3}+2^{n-2}-1.
\end{aligned}
\end{eqnarray}

\noindent{\bf{Proof of Theorem 4}}

Firstly we prove the following lemmas:

\noindent{\bf{Lemma}} {\bf{(1)}}:$\sum\limits_{i_1,\cdots,i_n=1}^2|\langle x^{i_1}_1\otimes x^{i_2}_2\otimes\cdots\otimes x^{i_n}_n\rangle^Q|\leq 2^{n-1}\sqrt{2}.$

\noindent{\bf{Lemma}} {\bf{(2)}}: $\sum\limits_{i_1,\cdots,i_n=1}^2|\langle x^{i_1}_1\otimes x^{i_2}_2\otimes\cdots\otimes x^{i_n}_n\rangle^Q|+\sum\limits_{i_1,\cdots, i_n\in C_0}|\langle x^{i_1}_1\otimes x^{i_2}_2\otimes\cdots\otimes x^{i_n}_n\rangle^Q|\leq 2^{n-1}\sqrt{3}+2^{n-1}-1$, where $C_0$ is the set consisting of all the strings with at least one $3,$
namely the set $\{30\cdots 0,030\cdots 0,\cdots, 00\cdots 03, 330\cdots,0,\cdots, 00\cdots,033,$ $\cdots, 33\cdots,3\}$. Here $\langle x \rangle^Q$ represents the expectation value
of $x$ with respect to the quantum states.

\medskip
\noindent Proof of \noindent{\bf{Lemma}} {\bf{(1)}}: Firstly we have
\begin{eqnarray}
\begin{aligned}
&|\langle x_1^1\rangle^Q|+|\langle x_1^2\rangle^Q|\\ \nonumber
\leq&\max\limits_{k_1,k_2} \{e_{k_1,k_2}\}=\max\limits_{k_1,k_2} \|T_{k_1,k_2}\|_2\leq \sqrt{2},
\end{aligned}
\end{eqnarray}
where $T_{k_1k_2}=(-1)^{k_1} x_1^1+(-1)^{k_2}x_1^2,$
$e_{k_1,k_2}$ is the maximum eigenvalue of $T_{k_1k_2},$ i.e., the $2$-norm of $T_{k_1k_2}.$  $e_{k_1,k_2}=\sqrt{2}$ for the mutually unbiased measurements $x_1^1$ and $x_2^2.$

From
\begin{eqnarray}
\begin{aligned}
&\sum\limits_{i_1,i_2=1}^2|\langle x_1^{i_1}\otimes x_2^{i_2}\rangle^Q|\\ \nonumber
\leq&\max\limits_{k_j,k_j',k_j^{''}} \|T_{k_1k_2}\otimes x_2^1+T_{k'_1k'_2}\otimes x_2^2\|\\ \nonumber
\leq &  \|T_{k_1k_2}\|_2\|x_2^1\|_2+\|T_{k'_1k'_2}\|_2\|x_2^2\|_2\\ \nonumber
\leq & \|T_{k_1k_2}\|_2+\|T_{k'_1k'_2}\|_2
\leq 2\sqrt{2},
\end{aligned}
\end{eqnarray}
where the first  inequality is attained by using the definition of $\sum\limits_{i_1,i_2=1}^2|\langle x_1^{i_1}\otimes x_2^{i_2}\rangle^Q|$, the second and the third inequalities are attained using the norm inequality and $\|x_k^{i_k}\|_2\leq 1,$
and the fourth inequality is due to $\|T_{k_1k_2}\|_2\leq \sqrt{2}$ and $\|T_{k'_1k'_2}\|_2\leq \sqrt{2}.$
Using the same method and the mathematical induction, we can prove the {\bf{Lemma}} {\bf{(1)}}.

\medskip
\noindent Proof of \noindent{\bf{Lemma}} {\bf{(2)}}: Firstly we have
\begin{eqnarray}
\begin{aligned}
&|\langle x_1^1\rangle^Q|+|\langle x_1^2\rangle^Q|+|\langle x_1^3\rangle^Q|\\ \nonumber
\leq&\max\limits_{k_1,k_2=0,1} \{e_{k_1,k_2}\}=\max\limits_{k_1,k_2=0,1} \|T_{k_1,k_2}\|_2\leq \sqrt{3},
\end{aligned}
\end{eqnarray}
where $T_{k_1k_2}=(-1)^{k_1} x_1^1+(-1)^{k_2} x_1^2+ x_1^3=[1,(-1)^{k_1}-(-1)^{k_2}\rm{i};(-1)^{k_1}+(-1)^{k_2}\rm{i},-1],$
$e_{k_1,k_2}$ is the maximum eigenvalue of $T_{k_1k_2}.$   $e_{k_1,k_2}=\sqrt{3}$ for the mutually unbiased measurements $x_1^1,$ $x_1^2$ and $x_2^3.$

Moreover,
\begin{eqnarray}
\begin{aligned}
&\sum\limits_{i_1,i_2=1}^2|\langle x_1^{i_1}\otimes x_2^{i_2}\rangle^Q|+\sum\limits_{i_1,i_2\in C_0}|\langle x_1^{i_1}\otimes x_2^{i_2}\rangle^Q|\\ \nonumber
\leq&\max\limits_{k_1,\cdots,k_5} e(T_{k_1\cdots k_5})+|\langle x_1^{3}\otimes x_2^{3}\rangle^Q|\\ \nonumber
\leq & 2\sqrt{3}+1,
\end{aligned}
\end{eqnarray}
where $T_{k_1\cdots k_5}=(-1)^{k_1}x_1^{1}\otimes x_2^{1}+(-1)^{k_2}x_1^{1}\otimes x_2^{2}+(-1)^{k_3}x_1^{2}\otimes x_2^{1}
+(-1)^{k_4}x_1^{2}\otimes x_2^{2}+(-1)^{k_5}x_1^{0}\otimes x_2^{3}+x_1^{3}\otimes x_2^{0}$ and $e(T_{k_1\cdots k_5})$ is the maximum eigenvalue of $T_{k_1\cdots k_5}.$
The first inequality is attained by using the definition of $|\langle x_1^{i_1}\otimes x_2^{i_2}\rangle^Q|$.
The maximum value of $e(T_{k_1\cdots k_5})$ is attained when $T_{k_1\cdots k_5}=[2,0,0,2\pm \rm{i};0,0,0,0;0,0,0,0;2\mp \rm{i},0,0,-2].$

Similarly we can prove the inequality for general $n$. $C_0$ can be split into two subsets $C_1$ and $C_2,$ where $C_1$ contains the elements with odd number of $3$ and $C_2=C'$ contains the elements with even number of $3.$
The maximum value of $\sum\limits_{i_1,\cdots,i_n\in C_2}|\langle x_1^{i_1}\otimes\cdots\otimes x_n^{i_n}\rangle^Q|$ is $2^{n-1}-1$ as each $|\langle x_1^{i_1}\otimes\cdots\otimes x_n^{i_n}\rangle^Q|\leq 1$, and the number of the elements in $C_2$ is $2^{n-1}-1.$  The maximum value of $\sum\limits_{i_1,\cdots,i_n=1}^2|\langle x_1^{i_1}\otimes\cdots\otimes x_n^{i_n}\rangle^Q|+\sum\limits_{i_1,\cdots,i_n\in C_1}|\langle x_1^{i_1}\otimes\cdots\otimes x_n^{i_n}\rangle^Q|\leq \max \rm{eig}(T)$ is attained when $T=\sum\limits_{i_1,\cdots,i_n=1}^2\pm x_1^{i_1}\otimes\cdots\otimes x_n^{i_n}+\sum\limits_{i_1,\cdots,i_n\in C_1}\pm x_1^{i_1}\otimes\cdots\otimes x_n^{i_n}$
is the matrix with entries $T_{11}=-T_{2^n,2^n}=2^{n-1},$ $T_{1,2^n}=2^{n-1}(1+\rm{i}),$  $T_{2^n,1}=2^{n-1}(1-\rm{i})$ and other entries being 0.  Here $\max \rm{eig}(T)=2^{n-1}\sqrt{3}$ is the maximum eigenvalue of $T$.
\smallskip
Next we prove the theorem. If $\rho_{A_1\cdots A_n B}$ admits LHV-BLHS model, we have
\begin{eqnarray}
\begin{aligned}
\label{star-n-2-c-e}
&\sum\limits_{i_1,\cdots,i_n\in C}|\langle x^{i_1}_1\otimes x^{i_2}_2\otimes \cdots \otimes x^{i_n}_n\otimes y_1^{i_1\cdots i_n}\rangle|^{\frac{1}{2}}\\
=&\sum\limits_{i_1,\cdots,i_n\in C}|\sum\limits_{\lambda_1\cdots\lambda_n}\sum\limits_{s=1}^{\lfloor\frac{n}{2}\rfloor}\sum\limits_{t=1}^{C_n^s}q_s^t p(\Lambda_s^t)p(\tilde{\Lambda}_s^t)\langle y^{i_1\cdots i_n}\rangle\\ \nonumber
&\times\langle x_{K_s^t}^{i_{K_s^t}}\rangle^Q_{\Lambda_s^t}\langle x_{\overline{K_s^t}}^{i_{\overline{K_s^t}}}\rangle^Q_{\tilde{\Lambda}_s^t} |^{\frac{1}{2}}\\ \nonumber
\leq &\sum\limits_{s=1}^{\lfloor\frac{n}{2}\rfloor}\sum\limits_{i_1,\cdots,i_n\in C}\sum\limits_{\lambda_1\cdots\lambda_n}\sum\limits_{t=1}^{C_n^s}|q_s^t p(\Lambda_s^t)p(\tilde{\Lambda}_s^t)\langle y^{i_1\cdots i_n}\rangle\\ \nonumber
\times&\langle x_{K_s^t}^{i_{K_s^t}}\rangle^Q_{\Lambda_s^t}\langle x_{\overline{K_s^t}}^{i_{\overline{K_s^t}}}\rangle^Q_{\tilde{\Lambda}_s^t} |^{\frac{1}{2}}\\ \nonumber
\end{aligned}
\end{eqnarray}
\begin{eqnarray}
\begin{aligned}
\leq &\sum\limits_{s=1}^{\lfloor\frac{n}{2}\rfloor}|\sum\limits_{t=1}^{C_n^s}\sum\limits_{\Lambda_s^t}\sqrt{q_s^t}p(\Lambda_s^t)2^{n-1-s}\langle x_{K_s^t}^{i_{K_s^t}}\rangle^Q_{\Lambda_{s}^t}|^{\frac{1}{2}}\\ \nonumber
&\times |\sum\limits_{t=1}^{C_n^s}\sum\limits_{\tilde{\Lambda}_k^s}\sqrt{q_s^t}p(\tilde{\Lambda}_s^t)\sum\limits_{i_{\overline{K_s^t}}}2^{s-1}\langle x_{\overline{K_s^t}}^{i_{\overline{K_s^t}}}\rangle^Q_{\tilde{\Lambda}_{s}^t}|^{\frac{1}{2}}\\ \nonumber
\leq& 2^{n-2}\sqrt{2},
\end{aligned}
\end{eqnarray}
where the first equality is attained by the definition of LHV-BLHS model, the first inequality is attained by $\sqrt{x_1+\cdots x_n}\leq \sqrt{x_1}+\cdots+\sqrt{x_n},$ the second inequality is attained by Cauchy-Schwarz inequality, and the last inequality is attained by the Lemma (1).

The inequality
$\sum\limits_{i_1,\cdots,i_n\in C\cup C'}|\langle x^{i_1}_1\otimes x^{i_2}_2\otimes \cdots \otimes x^{i_n}_n\otimes y_1^{i_1\cdots i_n}\rangle|^{\frac{1}{2}}
\leq 2^{n-2}\sqrt{3}+2^{n-2}-1$ can be similarly proved, by using the Cauchy-Schwarz inequality and Lemma (2).

\end{document}